\pdfoutput=1 

\documentclass[12pt,a4paper]{article}

\usepackage{epstopdf}
\usepackage{ifthen} 
\newboolean{pdflatex}
\setboolean{pdflatex}{true} 

\newboolean{articletitles}
\setboolean{articletitles}{true} 

\newboolean{uprightparticles}
\setboolean{uprightparticles}{false} 

\usepackage{microtype}
\usepackage{lineno}  
\usepackage{xspace} 

\usepackage{graphicx}  
\usepackage{color}
\usepackage{colortbl}
\graphicspath{{./figs/}} 

\usepackage{amsmath} 
\usepackage{amssymb}
\usepackage{amsfonts}
\usepackage{upgreek} 
\usepackage{array}
\newcommand*\patchAmsMathEnvironmentForLineno[1]{%
\expandafter\let\csname old#1\expandafter\endcsname\csname #1\endcsname
\expandafter\let\csname oldend#1\expandafter\endcsname\csname
end#1\endcsname
 \renewenvironment{#1}%
   {\linenomath\csname old#1\endcsname}%
   {\csname oldend#1\endcsname\endlinenomath}%
}
\newcommand*\patchBothAmsMathEnvironmentsForLineno[1]{%
  \patchAmsMathEnvironmentForLineno{#1}%
  \patchAmsMathEnvironmentForLineno{#1*}%
}
\AtBeginDocument{%
\patchBothAmsMathEnvironmentsForLineno{equation}%
\patchBothAmsMathEnvironmentsForLineno{align}%
\patchBothAmsMathEnvironmentsForLineno{flalign}%
\patchBothAmsMathEnvironmentsForLineno{alignat}%
\patchBothAmsMathEnvironmentsForLineno{gather}%
\patchBothAmsMathEnvironmentsForLineno{multline}%
}

\usepackage{hyperref}    
\usepackage[all]{hypcap} 

\def\lhcb {\mbox{LHCb}\xspace}
\def\ux85 {\mbox{UX85}\xspace}

\ifthenelse{\boolean{uprightparticles}}%
{
 
 \def\Pgamma      {\ensuremath{\upgamma}\xspace}

 \def\Ppi         {\ensuremath{\uppi}\xspace}

 \def\Ppsi        {\ensuremath{\uppsi}\xspace}

 \def\PDelta      {\ensuremath{\Delta}\xspace}                 
 \def\PXi      {\ensuremath{\Xi}\xspace}                 
 \def\PLambda      {\ensuremath{\Lambda}\xspace}                 
 \def\PSigma      {\ensuremath{\Sigma}\xspace}                 
 \def\POmega      {\ensuremath{\Omega}\xspace}                 
 \def\PUpsilon      {\ensuremath{\Upsilon}\xspace}                 
 
 \def\PB      {\ensuremath{\mathrm{B}}\xspace}                 
                  
 \def\PD      {\ensuremath{\mathrm{D}}\xspace}

 \def\PJ      {\ensuremath{\mathrm{J}}\xspace}                 
 \def\PK      {\ensuremath{\mathrm{K}}\xspace}

 \def\Pb      {\ensuremath{\mathrm{b}}\xspace}                 
 \def\Pc      {\ensuremath{\mathrm{c}}\xspace}

 \def\Pi      {\ensuremath{\mathrm{i}}\xspace}

 \def\Ps      {\ensuremath{\mathrm{s}}\xspace}

}
{
 
 \def\Pgamma      {\ensuremath{\gamma}\xspace}

 \def\Ppi         {\ensuremath{\pi}\xspace}

 \def\Ppsi        {\ensuremath{\psi}\xspace}                 
                  
 \mathchardef\PDelta="7101
 \mathchardef\PXi="7104
 \mathchardef\PLambda="7103
 \mathchardef\PSigma="7106
 \mathchardef\POmega="710A
 \mathchardef\PUpsilon="7107
                  
 \def\PB      {\ensuremath{B}\xspace}                 
                  
 \def\PD      {\ensuremath{D}\xspace}

 \def\PJ      {\ensuremath{J}\xspace}                 
 \def\PK      {\ensuremath{K}\xspace}

 \def\Pb      {\ensuremath{b}\xspace}                 
 \def\Pc      {\ensuremath{c}\xspace}

 \def\Pi      {\ensuremath{i}\xspace}

 \def\Ps      {\ensuremath{s}\xspace}

}

\def\g      {\ensuremath{\Pgamma}\xspace}

\def\squark    {\ensuremath{\Ps}\xspace}

\def\cquark    {\ensuremath{\Pc}\xspace}

\def\bquark    {\ensuremath{\Pb}\xspace}

\def\pion  {\ensuremath{\Ppi}\xspace}

\def\pip   {\ensuremath{\pion^+}\xspace}
\def\pim   {\ensuremath{\pion^-}\xspace}

\def\pipm  {\ensuremath{\pion^\pm}\xspace}
\def\pimp  {\ensuremath{\pion^\mp}\xspace}

\def\kaon  {\ensuremath{\PK}\xspace}
  \def\Kbar  {\kern 0.2em\overline{\kern -0.2em \PK}{}\xspace}

\def\Kz    {\ensuremath{\kaon^0}\xspace}
\def\Kzb   {\ensuremath{\Kbar^0}\xspace}
\def\KzKzb {\ensuremath{\Kz \kern -0.16em \Kzb}\xspace}
\def\Kp    {\ensuremath{\kaon^+}\xspace}
\def\Km    {\ensuremath{\kaon^-}\xspace}
\def\Kpm   {\ensuremath{\kaon^\pm}\xspace}
\def\Kmp   {\ensuremath{\kaon^\mp}\xspace}
\def\KpKm  {\ensuremath{\Kp \kern -0.16em \Km}\xspace}
\def\KS    {\ensuremath{\kaon^0_{\rm\scriptscriptstyle S}}\xspace}

  \def\Dbar    {\kern 0.2em\overline{\kern -0.2em \PD}{}\xspace}
\def\D       {\ensuremath{\PD}\xspace}
\def\Db      {\ensuremath{\Dbar}\xspace}
\def\Dz      {\ensuremath{\D^0}\xspace}
\def\Dzb     {\ensuremath{\Dbar^0}\xspace}
\def\DzDzb   {\ensuremath{\Dz {\kern -0.16em \Dzb}}\xspace}
\def\Dp      {\ensuremath{\D^+}\xspace}
\def\Dm      {\ensuremath{\D^-}\xspace}

\def\DpDm    {\ensuremath{\Dp {\kern -0.16em \Dm}}\xspace}

\def\Dstarp  {\ensuremath{\D^{*+}}\xspace}

\def\Dstarpm {\ensuremath{\D^{*\pm}}\xspace}

\def\B       {\ensuremath{\PB}\xspace}
  \def\Bbar    {\kern 0.18em\overline{\kern -0.18em \PB}{}\xspace}

\def\Bz      {\ensuremath{\B^0}\xspace}
\def\Bzb     {\ensuremath{\Bbar^0}\xspace}
\def\Bu      {\ensuremath{\B^+}\xspace}
\def\Bub     {\ensuremath{\B^-}\xspace}
\def\Bp      {\ensuremath{\Bu}\xspace}
\def\Bm      {\ensuremath{\Bub}\xspace}
\def\Bpm     {\ensuremath{\B^\pm}\xspace}

\def\Bs      {\ensuremath{\B^0_\squark}\xspace}

\def\jpsi     {\ensuremath{{\PJ\mskip -3mu/\mskip -2mu\Ppsi\mskip 2mu}}\xspace}
\def\psitwos  {\ensuremath{\Ppsi{(2S)}}\xspace}

  \def\Y#1S{\ensuremath{\PUpsilon{(#1S)}}\xspace}

\def\L {\ensuremath{\PLambda}\xspace}
\def\Lbar {\ensuremath{\kern 0.1em\overline{\kern -0.1em\PLambda}}\xspace}

\def\Lb      {\ensuremath{\L^0_\bquark}\xspace}

\def\Lc      {\ensuremath{\L^+_\cquark}\xspace}

\def\to                 {\ensuremath{\rightarrow}\xspace}

\def\CP    {\ensuremath{C\!P}\xspace}

\def\AT#1     {\ensuremath{A_{\mathrm{T}}^{#1}}\xspace}           

\def\C#1      {\ensuremath{\mathcal{C}_{#1}}\xspace}                       
\def\Cp#1     {\ensuremath{\mathcal{C}_{#1}^{'}}\xspace}                    
\def\Ceff#1   {\ensuremath{\mathcal{C}_{#1}^{\mathrm{(eff)}}}\xspace}        
\def\Cpeff#1  {\ensuremath{\mathcal{C}_{#1}^{'\mathrm{(eff)}}}\xspace}       
\def\Ope#1    {\ensuremath{\mathcal{O}_{#1}}\xspace}                       
\def\Opep#1   {\ensuremath{\mathcal{O}_{#1}^{'}}\xspace}                    

\newcommand{\ket}[1]{\ensuremath{|#1\rangle}}              

\newcommand{\tev}{\ensuremath{\mathrm{\,Te\kern -0.1em V}}\xspace}
\newcommand{\gev}{\ensuremath{\mathrm{\,Ge\kern -0.1em V}}\xspace}
\newcommand{\mev}{\ensuremath{\mathrm{\,Me\kern -0.1em V}}\xspace}
\newcommand{\kev}{\ensuremath{\mathrm{\,ke\kern -0.1em V}}\xspace}
\newcommand{\ev}{\ensuremath{\mathrm{\,e\kern -0.1em V}}\xspace}
\newcommand{\gevc}{\ensuremath{{\mathrm{\,Ge\kern -0.1em V\!/}c}}\xspace}
\newcommand{\mevc}{\ensuremath{{\mathrm{\,Me\kern -0.1em V\!/}c}}\xspace}
\newcommand{\gevcc}{\ensuremath{{\mathrm{\,Ge\kern -0.1em V\!/}c^2}}\xspace}
\newcommand{\gevgevcccc}{\ensuremath{{\mathrm{\,Ge\kern -0.1em V^2\!/}c^4}}\xspace}
\newcommand{\mevcc}{\ensuremath{{\mathrm{\,Me\kern -0.1em V\!/}c^2}}\xspace}

\def\invfb   {\ensuremath{\mbox{\,fb}^{-1}}\xspace}

\def\gsim{{~\raise.15em\hbox{$>$}\kern-.85em
          \lower.35em\hbox{$\sim$}~}\xspace}
\def\lsim{{~\raise.15em\hbox{$<$}\kern-.85em
          \lower.35em\hbox{$\sim$}~}\xspace}

\def\ptot       {\mbox{$p$}\xspace}
\def\pt         {\mbox{$p_{\rm T}$}\xspace}

\def\evtgen     {\mbox{\textsc{EvtGen}}\xspace}
\def\pythia     {\mbox{\textsc{Pythia}}\xspace}

\def\geant      {\mbox{\textsc{Geant4}}\xspace}

\def\photos     {\mbox{\textsc{Photos}}\xspace}

\def\tell1  {TELL1\xspace}
\def\ukl1   {UKL1\xspace}

\def\ads {{\rm ADS}\xspace}

\newcommand{\bea}{\begin{eqnarray}}
\newcommand{\eea}{\end{eqnarray}}
\newcommand{\beq}{\begin{equation}}
\newcommand{\eeq}{\end{equation}}      

\usepackage{multirow}

\usepackage{cite} 
\usepackage{mciteplus}

\usepackage{comment}

\usepackage{enumerate}

\usepackage[font={small}]{caption}
\begin{document}

\renewcommand{\thefootnote}{\fnsymbol{footnote}}
\setcounter{footnote}{1}


\begin{titlepage}
\pagenumbering{roman}

\vspace*{-1.5cm}
\centerline{\large EUROPEAN ORGANIZATION FOR NUCLEAR RESEARCH (CERN)}
\vspace*{1.5cm}
\hspace*{-0.5cm}
\begin{tabular*}{\linewidth}{lc@{\extracolsep{\fill}}r}
\vspace*{-1.5cm}\mbox{\!\!\!\includegraphics[width=.12\textwidth]{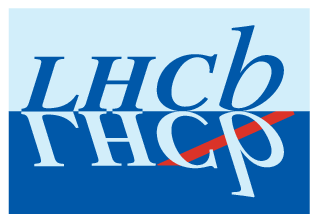}} & &
\\
 & & CERN-PH-EP-2013-038 \\  
 & & LHCb-PAPER-2012-055 \\  
& & May 4, 2013 \\ 
& & \\
\end{tabular*}

\vspace*{2.5cm}

{\bf\boldmath\huge
\begin{center}
 Observation of the \\ suppressed ADS modes $B^\pm \to [\pi^\pm K^\mp \pi^+\pi^-]_D K^\pm$ and  
$B^\pm \to [\pi^\pm K^\mp \pi^+\pi^-]_D \pi^\pm$ 
\end{center}
}

\vspace*{1.5cm}

\begin{center}
The LHCb collaboration\footnote{Authors are listed on the following pages.}
\end{center}

\vspace*{0.5cm}

\begin{abstract}
\vspace*{0.2cm}
  \noindent
An analysis of $B^{\pm}\to DK^{\pm}$ and $B^{\pm}\to D\pi^{\pm}$ decays is presented where the $D$ meson is reconstructed in the four-body final state $K^{\pm}\pi^{\mp} \pi^+ \pi^-$. Using LHCb data corresponding to an integrated luminosity of $1.0{\rm \,fb}^{-1}$, first observations are made of the suppressed ADS modes $B^{\pm}\to [\pi^{\pm} K^{\mp}\pi^+\pi^-]_D K^{\pm}$ and $B^{\pm}\to [\pi^{\pm} K^{\mp} \pi^+\pi^- ]_D\pi^{\pm}$ with a significance of $5.1\sigma$ and greater than $10\sigma$, respectively. Measurements of \CP asymmetries and \CP-conserving ratios of partial widths from this family of decays are also performed.   The magnitude of the ratio between the suppressed and favoured $B^{\pm}\to DK^{\pm}$ amplitudes is determined to be $r^K_B = 0.097 \pm{0.011}$.
\end{abstract}

\vspace*{1.0cm}

\begin{center}
Submitted to Phys. Lett. B
\end{center}

\vspace{\fill}

{\footnotesize 
\centerline{\copyright~CERN on behalf of the \lhcb collaboration, license \href{http://creativecommons.org/licenses/by/3.0/}{CC-BY-3.0}.}}
\vspace*{2mm}

\end{titlepage}


\newpage
\setcounter{page}{2}
\mbox{~}
\newpage

\centerline{\large\bf LHCb collaboration}
\begin{flushleft}
\small
R.~Aaij$^{40}$, 
C.~Abellan~Beteta$^{35,n}$, 
B.~Adeva$^{36}$, 
M.~Adinolfi$^{45}$, 
C.~Adrover$^{6}$, 
A.~Affolder$^{51}$, 
Z.~Ajaltouni$^{5}$, 
J.~Albrecht$^{9}$, 
F.~Alessio$^{37}$, 
M.~Alexander$^{50}$
% ...

{\footnotesize \it
% ...
$ ^{5}$Clermont Universit\'{e}, Universit\'{e} Blaise Pascal, CNRS/IN2P3, LPC, Clermont-Ferrand, France\\
$ ^{6}$CPPM, Aix-Marseille Universit\'{e}, CNRS/IN2P3, Marseille, France\\
% ...
$ ^{9}$Fakult\"{a}t Physik, Technische Universit\"{a}t Dortmund, Dortmund, Germany\\
% ...
$ ^{35}$Universitat de Barcelona, Barcelona, Spain\\
$ ^{36}$Universidad de Santiago de Compostela, Santiago de Compostela, Spain\\
$ ^{37}$European Organization for Nuclear Research (CERN), Geneva, Switzerland\\
% ...
$ ^{40}$Nikhef National Institute for Subatomic Physics, Amsterdam, The Netherlands\\
% ...
$ ^{45}$H.H. Wills Physics Laboratory, University of Bristol, Bristol, United Kingdom\\
% ...
$ ^{50}$School of Physics and Astronomy, University of Glasgow, Glasgow, United Kingdom\\
$ ^{51}$Oliver Lodge Laboratory, University of Liverpool, Liverpool, United Kingdom\\
% ...
$ ^{n}$LIFAELS, La Salle, Universitat Ramon Llull, Barcelona, Spain\\
% ...
}
\end{flushleft}

\cleardoublepage


\renewcommand{\thefootnote}{\arabic{footnote}}
\setcounter{footnote}{0}



\pagestyle{plain} 
\setcounter{page}{1}
\pagenumbering{arabic}


%

\section{Introduction}
\label{sec:intro}

A measurement of the angle \g  (also denoted as $\phi_3$) of the CKM Unitarity Triangle~\cite{Cabibbo:1963yz,*Kobayashi:1973fv}  in processes involving tree-level decays provides a Standard Model (SM) benchmark against which observables more sensitive to new physics contributions can be compared. Currently such comparisons are limited by the relatively large uncertainty ($\sim 12^\circ$~\cite{PhysRevD.84.033005, *Bona:2005vz}) on the determination of \g in tree-level decays~\cite{Lees:2013zd,Trabelsi:2013uj}. More precise measurements are therefore required.

A powerful strategy to measure \g in tree-level processes is to study \CP-violating observables in the decays $B^\pm \to \D h^\pm$,
where \D indicates a neutral charm meson which decays in a  mode common to both \Dz and \Dzb states, and $h$, the bachelor hadron, is either a kaon or a pion. In the case of $B^{-} \to \D K^{-}$, interference occurs between the suppressed $b \to u \bar{c} s$ and favoured $b \to c \bar{u} s$ decay paths, and similarly for the charge conjugate decay. The magnitude of the interference is governed by three parameters: the weak-phase difference, \g, the \CP-conserving strong-phase difference, $\delta^K_B$, and the ratio of the magnitudes of the two amplitudes, $r^K_B$. Similar interference effects occur in the case when the bachelor hadron is a pion, but additional Cabibbo suppression factors mean that the sensitivity to \g is much diluted. Many possibilities exist for the \D decay mode, including \CP eigenstates~\cite{Gronau:1990ra,*Gronau:1991dp} and  self-conjugate three-body decays~\cite{GGSZ1,*GGSZ2}, which have both been exploited by LHCb in recent measurements~\cite{LHCBADSGLW2BODY,LHCBGGSZMODIND}. Results of LHCb have also been presented making use of a similar strategy with $\Bz/\Bzb$ mesons~\cite{BIGED}. Another option, termed the `ADS' method in reference to its originators~\cite{Atwood:1996ci,*Atwood:2000ck}, is to consider modes such as $\D \to K^{\pm} \pi^{\mp}$ and to focus on the suppressed final-state $B^\pm \to [\pi^\pm K^\mp]_{D} K^\pm$, in which the favoured $B^\pm$ decay is followed by a doubly Cabibbo-suppressed \D decay, or the suppressed $B^\pm$ decay precedes a favoured \D decay. The amplitudes of such combinations are of similar total magnitude and hence large interference can occur, giving rise to significant \CP-violating effects. In contrast, the interference in the favoured decay $B^\pm \to [K^\pm \pi^\mp]_D K^\pm$ is low.

In this Letter, a search is performed for the previously unobserved ADS decays $B^\pm \to [\pi^\pm K^\mp \pi^+ \pi^-]_{D} h^{\pm}$. The $D$ decay is treated inclusively, with no attempt to separate out the intermediate resonances contributing to the four-body final state. LHCb  has already presented an ADS study using  $B^\pm \to [\pi^{\pm} K^{\mp}]_{D} h^\pm$ decays, and many features of the current analysis are similar to those of the earlier paper~\cite{LHCBADSGLW2BODY}. In this study a total of seven observables is measured: the ratio of partial widths involving the favoured modes
\beq
R_{K/\pi}^{K3\pi} \equiv \frac{ \Gamma(\Bm\to [\Km\pi^+\pi^-\pi^+]_D\Km)+\Gamma(\Bp\to [\Kp \pi^-\pi^+\pi^-]_D\Kp) }
{ \Gamma(\Bm\to [\Km\pi^+\pi^-\pi^+]_D\pim)+\Gamma(\Bp\to [\Kp \pi^-\pi^+\pi^- ]_D\pip) },
\eeq
\noindent
two \CP asymmetries, again involving the favoured modes
\beq
A_{h}^{K3\pi} \equiv \frac{ \Gamma(\Bm\to [\Km\pi^+\pi^+\pi^-]_Dh^-)-\Gamma(\Bp\to [\Kp \pi^-\pi^+\pi^-]_Dh^+) }{ \Gamma(\Bm\to [ \Km\pi^+\pi^+\pi^-]_Dh^-)+\Gamma(\Bp\to [\Kp \pi^-\pi^+\pi^- ]_Dh^+) },
\eeq
\noindent
and four charge-separated partial widths of the suppressed \ads mode relative to the favoured mode
\beq
R_h^{K3\pi, \pm}  \equiv \frac{ \Gamma(\Bpm\to [\pipm \Kmp \pi^+ \pi^-]_Dh^{\pm})}{ \Gamma(\Bpm\to  [\Kpm \pimp \pi^+\pi^-]_Dh^{\pm})}.
\eeq

The observables  $R_h^{K3\pi, \pm}$ carry the highest sensitivity to $\gamma$, $r_B^h$ and $\delta_B^h$. They are related by the expression~\cite{Atwood:1996ci,*Atwood:2000ck,COHERENCE}
\beq
R_h^{K3\pi, \pm} = {(r^h_B)^2 + (r_D^{K3\pi})^2 + 2 r^h_B r^{K3\pi}_D \kappa_D^{K3\pi} \cos (\delta_B^h + \delta_D^{K3\pi} \pm \gamma}).
\label{eq:rads}
\eeq
Here  $r_D^{K3\pi}$ is the ratio of the magnitudes of the doubly Cabibbo-suppressed and Cabibbo-favoured \D decay amplitudes,
and  $\delta_D^{K3\pi}$ is the strong-phase difference between the amplitudes, averaged over the final-state phase space.
The coherence factor $\kappa_D^{K3\pi}$ 
accounts for possible dilution effects in the interference arising from the contribution of several intermediate resonances in the $D$ decay~\cite{COHERENCE}.
Information on these $D$ decay parameters is available from external sources.  
Branching ratio measurements indicate that $r^{K3\pi}_D\sim 0.06$~\cite{PDG2012}.  Studies of  
quantum-correlated $\D\Db$ pairs, performed at CLEO-c,  yield $\delta_D^{K3\pi} = (114 \, ^{+26}_{-23})^\circ$
and $\kappa_D^{K3\pi} = 0.33 \, ^{+0.20}_{-0.23}$~\cite{NORMLOWREY}.\footnote{The phase $\delta_D^{K3\pi}$ is given in the convention where $\CP \ket{\Dz} = \ket{\Dzb}$.}
The relatively low value of the coherence factor limits the sensitivity of  $R_h^{K3\pi, \pm}$ to $\gamma$ and $\delta_B^h$,  but does not hinder this observable in providing information on $r_B^h$.   Improved knowledge of $r_B^h$ is valuable in providing a constraint which other $B^\pm \to D h^\pm$ analyses can benefit from.

\section{The \lhcb detector and the analysis sample}
\label{sec:dataset}

This analysis uses data, corresponding to an integrated luminosity of 1.0~\invfb, collected by \lhcb in 2011 at $\sqrt{s}=7~\tev$.
The \lhcb experiment~\cite{Alves:2008zz} takes advantage of the high $b\bar{b}$ and $c\bar{c}$ production cross sections at the Large Hadron Collider to record large samples of heavy-hadron decays.
It instruments the pseudorapidity range $2<\eta <5$ of the proton-proton ($pp$) collisions with a dipole magnet and a tracking system that achieves a momentum resolution of $0.4-0.6\%$ in the range $5-100$~\gevc. 
The dipole magnet can be operated in either polarity and this feature is used to reduce systematic effects due to detector asymmetries. In 2011, 58\% of the data were taken with one polarity, 42\% with the other.
The $pp$ collisions take place inside a silicon microstrip vertex detector that provides clear separation of secondary $B^\pm$ vertices from the primary collision vertex (PV) as well as discrimination for tertiary $D$ vertices.
Two ring-imaging Cherenkov (RICH) detectors~\cite{Adinolfi:2012an} with three radiators (aerogel, ${\rm C_{4}F_{10}}$ and ${\rm CF_4}$) provide dedicated particle identification (PID), which is critical for the separation of $\B^\pm\! \to\! D K^\pm$ and $\B^\pm \! \to\! D \pi^\pm$ decays.

A two-stage trigger is employed. First, a hardware-based decision is taken at a rate of up to 40~MHz. 
It accepts high transverse energy clusters in either an electromagnetic  or hadron calorimeter, or a muon of high momentum transverse to the beam line (\pt).
For this analysis, it is required that either one of the five tracks forming the $B^\pm$ candidate points at a cluster in the hadron calorimeter, or that the hardware-trigger decision was taken independently of any of these tracks.
A subsequent trigger level, implemented entirely in software, receives events at a rate of 1 MHz and retains $\sim0.3\%$ of them.
At least one track should have $\pt > 1.7$ \gevc and impact parameter (IP) $\chi^2$ with respect to the PV greater than 16. The IP $\chi^2$ is defined as the difference between the $\chi^2$ of the PV reconstructed with and without the considered track. In order to maximise efficiency at an acceptable trigger rate, a displaced vertex is selected with a decision tree algorithm that uses flight distance as well as fit quality, \pt and information on the IP  with respect to the PV of the tracks.  More information can be found in Ref.~\cite{LHCbTrigger}. Full event reconstruction occurs offline, and a loose selection is run to reduce the size of the sample prior to final analysis.  This selection consists of a decision tree algorithm similar to that used in the trigger, but in this case the entire decay chain is fully reconstructed and the selection benefits from the improved quality of the offline reconstruction.

Approximately one million simulated signal events are used in the analysis as well as a sample of  $\sim 10^8$ generic $B_{q}\to DX$ decays, where $q\in\{u,d,s\}$. These samples are generated using \pythia6.4~\cite{Sjostrand:2006za} configured with parameters detailed in Ref.~\cite{LHCBTUNE}. The \evtgen package~\cite{Lange:2001uf} is used to generate hadronic decays, in which final state radiation is generated using the \photos package~\cite{PHOTOS}. The interaction of the generated particles with the LHCb detector is simulated using the \geant toolkit~\cite{Allison:2006ve, *Agostinelli:2002hh} as described in Ref.~\cite{CLEMENCIC}.

\section{Candidate selection and background rejection}
\label{sec:select}

The reconstruction considers all  $\B^\pm \to \D h^\pm$ channels of interest. The reconstructed \D candidate mass is required to be within  $\pm25\mevcc$ ($\approx 3.5 \sigma$) of its nominal value~\cite{PDG2012}. The \D daughter tracks are required to have $\pt > 0.25$\gevc, while the bachelor track is required to satisfy $0.5 < \pt < 10$\gevc and $5 < p < 100$\gevc. The tighter requirements on the bachelor track ensure that it resides within the kinematic coverage of the PID calibration samples acquired through the decay mode $\Dstarp\to \Dz \pip, \Dz \to\Km\pip$. Details of the PID calibration procedure are given in Sect.~\ref{sec:yield}. Furthermore, a kinematic fit is performed to each decay chain~\cite{2005NIMPA.552..566H} constraining both the \Bpm and \D vertices to points in 3D space, while simultaneously constraining the \D candidate to its nominal mass.  This fit results in a \Bpm mass resolution of 15\mevcc, a 10\% improvement with respect to the value prior to the fit. Candidates are retained that have an invariant mass in the interval $5120 - 5750$\mevcc.

A boosted decision tree (BDT) discriminator~\cite{Breiman}, implementing the \emph{GradientBoost} algorithm~\cite{Friedman2002367}, is employed to achieve further background suppression. The BDT is trained using the simulated $\B^\pm \to \D h^\pm$ events together with a pure sample of combinatoric background candidates taken from a subset of the data in the invariant mass range $5500 - 5800$\mevcc. The BDT considers a variety of properties associated with each signal candidate. These properties can be divided into two categories; (i) quantities common to both the tracks and to the \D and \Bpm candidates, (ii) quantities associated with only the \D and \Bpm candidates. Specifically, the properties considered in each case are as follows:
\begin{enumerate}[(i)]
\item \ptot, \pt and IP $\chi^2$;
\item decay time, flight distance from the PV, vertex quality, and the angle between the particle's momentum vector and a line connecting the PV to the particle's decay vertex.
\end{enumerate}
In addition, the BDT also considers information from the rest of the event through an isolation variable that represents the imbalance of \pt around the \Bpm candidate. The variable is defined as 
\begin{equation}
A_{\pt}=\frac{\pt(\Bpm)-\sum\pt}{\pt(\Bpm)+\sum\pt},
\end{equation}
where $\sum\pt$ corresponds to the sum of \pt over all tracks identified, within a cone of half-angle 1.5  in pseudorapidity and 1.5~rad in azimuthal angle, that are not associated with the signal \Bpm candidate. Since no PID information is used as input during the training, the BDT has very similar performance for both $\B^\pm\to \D K^\pm$ and $\B^\pm\to \D \pi^\pm$ decay modes, with small differences arising from the variation in kinematics between the two. 

The optimal cut value on the BDT response is chosen by considering the combinatorial background level ($b$) in the invariant mass distribution of the favoured $\Bpm\to \D \pipm$ final state. The large signal peak in this sample is scaled by the predicted branching fraction of the suppressed mode for the case when the interference amongst the intermediate resonances of the \D decay is maximally destructive. This provides a conservative estimate of the suppressed-sign signal yield ($s$). It is then possible to construct the quantity $s/\sqrt{s + b}$ to serve as an optimisation metric. Assessment of this metric finds an optimal working point where a signal efficiency of $\sim 85\%$ is expected while rejecting $> 99\%$ of the combinatorial background. This same working point is used in selecting both suppressed and favoured final states.

PID information is quantified as differences between the logarithm of likelihoods, $\ln\mathcal{L}_h$, under five mass hypotheses, $h\in\{\pi,K,p,e,\mu\}$ (DLL). The daughter kaon from the \D meson decay is required to satisfy ${\rm DLL}_{K\pi} \equiv\ln\mathcal{L}_{K}-\ln\mathcal{L}_{\pi}>2$, while the daughter pions must have ${\rm DLL}_{K\pi}<-2$.  A sample enriched in $\B^\pm \to \D K^\pm$ decays is selected by requiring ${\rm DLL}_{K\pi}>4$ for the bachelor hadron.  Candidates failing this cut are retained in a separate sample, which is predominantly composed of $\B^\pm \to \D \pi^\pm$ decays.

Backgrounds from genuine \Bpm decays that do not involve a true $\D$ meson are suppressed by requiring the flight distance significance of the \D candidate from the \Bpm vertex  be greater than $2$. The branching ratios of five-body charmless decays are currently unmeasured, and so the residual contamination from this source is estimated by assuming that the proportion of these decays passing the $\B^\pm \to [\pi^\pm K^\mp \pi^+\pi^-]_D  h^\pm$ selection is the same as the proportion of three-body charmless decays passing the  $B^\pm \to [\pi^\pm K^\mp ]_D  h^\pm$ selection reported in Ref.~\cite{LHCBADSGLW2BODY}. In the case of $\Bpm \to \pi^+\pi^-\pi^+\pi^- K^\pm$ this assumption can be validated by removing the flight significance cut and inspecting the sideband above the $D$ mass after adjusting the selection to isolate $\B^\pm \to [\pi^+  \pi^- \pi^+\pi^-]_D  K^\pm$ decays. A $\Bpm \to \pi^+\pi^-\pi^+\pi^- K^\pm$ signal is observed with a magnitude compatible with that found when scaling the results of the analogous exercise performed with a  $\B^\pm \to [\pi^+\pi^-]_D  K^\pm$ selection, and is eliminated when the flight-significance cut is reinstated. 
Following these studies the residual contamination from five-body charmless decays is determined to be $2.2 \pm 1.1$ candidates for the suppressed $\Bpm \to D K^\pm$ selection, and negligible for all other samples.

Contamination involving misidentified charmonium decays is eliminated by considering the possible neutral combinations of the bachelor track and any one of the \D daughter tracks under the hypothesis that both tracks are muons. For those combinations where both tracks satisfy a loose muon PID requirement, the parent \Bpm candidate is vetoed if the invariant mass of this combination is within $\pm22$\mevcc of either the \jpsi or \psitwos mass~\cite{PDG2012}.

The suppressed signal sample suffers a potentially large cross-feed from favoured signal decays in which a $K^\pm$ and  $\pi^\mp$ from the \D decay are misidentified as  $\pi^\pm$ and  $K^\mp$, respectively.  This contamination is reduced by vetoing any suppressed  candidate whose reconstructed \D mass, under the exchange of mass hypotheses between the daughter kaon and either of the two same-sign daughter pions, lies within $\pm15$\mevcc of the nominal \D mass. For the measurement of $R_h^{K3\pi,\pm}$  this veto is also applied to the favoured  mode.   Study of the cross-feed contamination in the mass sidebands of the \D candidates allows the estimate of the residual contamination in the signal region to be checked. 
The residual cross-feed after all selection requirements is estimated to be  $(7.1 \pm 3.1)\times10^{-5}$. 

The mass window for the \D candidates is sufficiently tight to eliminate background arising from
single-track misidentifications in  the four-body decays $\D \to K^+K^+ \pi^+\pi^-$ and $\D \to \pi^+\pi^-\pi^+\pi^-$. The good performance of the RICH system ensures that the residual background from  $\D \to K^+K^-\pi^+\pi^-$ decays in which three tracks are misidentified is negligible.  The contamination in the suppressed   $\B^\pm\to  [\pi^\pm K^\mp \pi^+\pi^-]_D  K^\pm$ sample from  $\Bpm \to [\Kmp \KS \pim]_{D} \Kpm$ with $\KS \to \pip\pim$ is estimated from simulation to occur at the rate of $(6.1 \pm 1.7) \times 10^{-5}$.

Only one candidate per event is retained for analysis.  In the $0.8\%$ of events that contain more than one candidate a choice is made by selecting the candidate with the best-quality \Bpm vertex.

Using simulation it is found that the selection leads to an acceptance across the four-body phase space of the $D$ decay  that is uniform to a good approximation. This property is important as it means that the values of the coherence factor and strong-phase difference measured in Ref.~\cite{NORMLOWREY}, which are integrated over all phase space,  can be applied when interpreting the results of the current analysis. It is verified using simulation that the small non-uniformities that exist in the acceptance induce negligible bias in the effective value of these parameters.

\section{Signal yields and systematic uncertainties}
\label{sec:yield}

The observables of interest are determined with a binned maximum-likelihood fit to the invariant mass distributions of the selected \Bpm candidates.
Distinguishing between \Bp and  \Bm candidates, favoured and suppressed decay topologies, and those that pass or fail the bachelor PID requirement imposed to select $\Bpm \to D K^\pm$ decays, yields eight disjoint subsamples, which are fitted simultaneously.
The total probability density function (PDF) used in the fit is built from four main sources representing the various categories of candidates in each subsample.

\begin{enumerate}
\item {\boldmath {$\Bpm\to \D \pipm$}}\\
In the subsamples failing the bachelor PID cut, a modified Gaussian function,
\begin{equation}
f(m) \propto \exp\left(\frac{-(m-\mu)^2}{2\sigma^2+(m-\mu)^2\alpha_{L,R}}\right) \label{eq:tails}
\end{equation}
describes the asymmetric function of peak of value $\mu$ and width $\sigma$ where $\alpha_L(m<\mu)$ and $\alpha_R(m>\mu)$ parameterise the tails. True $\Bpm\to \D \pipm$ candidates that pass the PID cut are reconstructed as $\Bpm\to \D\Kpm$.
As these candidates have an incorrect mass assignment they form a displaced mass peak with a tail that extends to higher invariant mass.
These candidates are modelled by the sum of two Gaussian PDFs, also altered to include tail components as in Eq.~\ref{eq:tails}.
All parameters are allowed to vary except the lower-mass tail which is fixed to the value found in simulation to ensure fit stability, and later considered amongst the systematic uncertainties. These shapes are considered identical for \Bp and \Bm decays. 
\item {\boldmath {$\Bpm\to \D \Kpm$}}\\
In the subsamples that pass the PID cut on the bachelor, the same modified Gaussian function as quoted in Eq.~\ref{eq:tails} is used.
The peak value and the two tail parameters are identical to those of the higher $\Bpm\to \D \pipm$ peak.
The width is $0.95\pm0.02$ times the $\Bpm \to \D \pipm$ width, as determined by a separate study of the favoured mode. Candidates failing the PID cut are described by a fixed shape that is obtained from simulation and later varied to assess the systematic uncertainty.
\item {\bf Partially reconstructed {\boldmath ${b}$}-hadron decays}\\
Partially reconstructed decays populate the invariant mass region below the \Bpm mass. Such candidates may enter the signal region, especially where Cabibbo-favoured $\Bpm \to X \D\pipm$ modes are misidentified as $\Bpm \to \D\Kpm$. The large simulated sample of inclusive $\B_q \to DX$ decays is used to model this background. After applying the selection, two non-parametric PDFs~\cite{Cranmer:2000du} are defined (for the $\D\pipm$ and $\D \Kpm$ selections) and used in the signal  fit for both the favoured and suppressed mode subsamples. 

In addition, partially reconstructed $\Bs \to D \Km\pip$  and $\Lb \to [p\Km\pip\pim\pip]_{\Lc} h^{-}$  decays and their charge-conjugated modes are considered as background sources specific to the  suppressed $\Bpm \to \D \Kpm$ and favoured mode subsamples, respectively.
PDFs for both these sources of background are determined from simulation and smeared to match the resolution observed in data. 

The yield of these background components  in each subsample varies independently in the fit, making no assumption of \CP symmetry.
\item {\bf Combinatoric background} \\ A linear approximation is adequate to describe the distribution across the invariant mass spectrum considered. A common shape parameter is used in all subsamples, though yields vary independently.

\end{enumerate}

The proportion of $\Bpm \to \D h^{\pm}$ passing or failing the PID requirement is determined from an analysis of approximately 20~million \Dstarpm decays reconstructed as $\Dstarpm\to D \pipm,\,  D \to\Kmp\pipm$.   The reconstruction is performed using only kinematic variables, and provides a high purity calibration sample of $K$ and $\pi$ tracks which is unbiased for studies exploiting the RICH and is therefore made use of to measure the 
PID efficiency as a function of track momentum, pseudorapidity and number of tracks in the detector. Through reweighting the calibration spectra in these variables to match that of the candidates in the $\Bpm\to \D \pipm$ peak, the effective PID efficiency of the signal is determined. This data-driven approach finds a retention rate on the bachelor track of 86.1\% and 3.7\% for kaons and pions, respectively.
An absolute $1.0\%$ systematic uncertainty on the kaon efficiency is estimated from simulation. The $\Bpm \to \D \pipm$ fit to data becomes significantly incorrect when the PID efficiency is varied outside the absolute range of $\pm0.2\%$, and so this value is taken as the systematic uncertainty for pions.

\label{sec:syst}
Detection and production asymmetries are accounted for using the same procedure followed in Ref.~\cite{LHCBADSGLW2BODY}, based on the measurement of the observed raw asymmetry of  $\Bpm\to\jpsi\Kpm$ decays in the \lhcb detector~\cite{Aaij:2012jw}.
A detection asymmetry of $(-0.5 \pm 0.7)$\% is assigned for each unit of strangeness in the final state to account for the different interaction lengths of \Km and \Kp mesons. The equivalent asymmetry for pions is expected to be much smaller and ($ 0.0 \pm 0.7$)\% is assigned. This uncertainty also accounts for the residual physical asymmetry between the left and right sides of the detector after summing both magnet-polarity data sets. Simulation of $b$-hadron production in $pp$ collisions suggests a small excess of \Bp over \Bm mesons. 
A production asymmetry of $(-0.8\pm0.7)$\% is assumed in the fit such that the combination of these estimates aligns with the observed raw asymmetry of $\Bpm\to\jpsi\Kpm$ decays~\cite{Aaij:2012jw}. 

\begin{figure*}[htb]
\begin{center}
\includegraphics[width=0.99\textwidth]{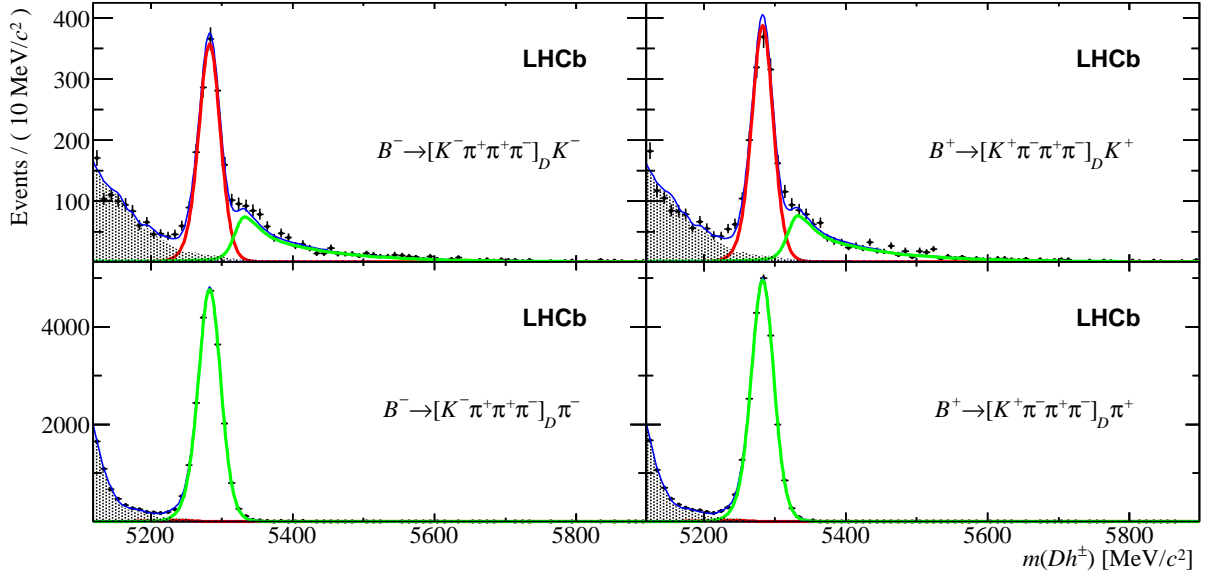}
\caption{Invariant mass distributions of selected $\Bpm\to[\Kpm\pi^{\mp}\pip\pim]_Dh^{\pm}$ candidates, separated by charge. The left plots are \Bm candidates, \Bp are on the right. In the top plots, the bachelor track passes the PID cut and the $\Bpm$ candidates are reconstructed assigning this track the kaon mass. The remaining candidates are placed in the sample displayed on the bottom row and are reconstructed with a pion mass hypothesis. The dark (red) and light (green) curves represent the fitted $\Bpm \to \D \Kpm$ and $\Bpm \to \D \pipm$ components, respectively. The shaded contribution indicates partially reconstructed decays and the total PDF includes the combinatorial component.
\label{fig:d2kpipipi}}
\end{center}
\end{figure*}

\begin{figure*}[htb]
\begin{center}
\includegraphics[width=0.99\textwidth]{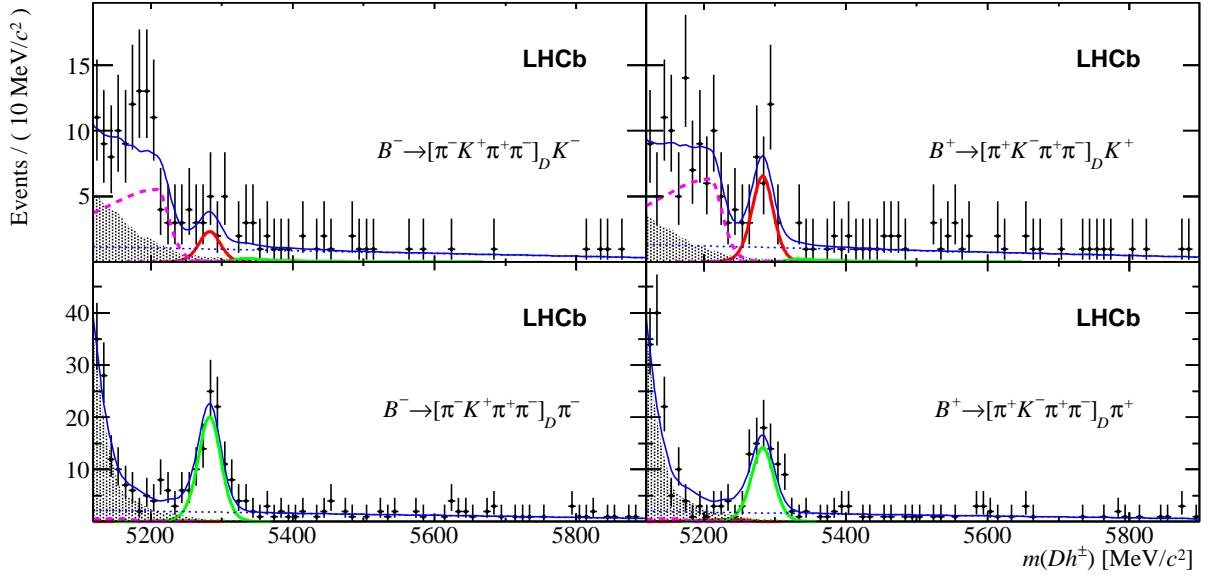}
\caption{Invariant mass distributions of selected $\Bpm\to[\pipm K^{\mp} \pip \pim]_Dh^{\pm}$ decays, separated by charge. See the caption of Fig.~\ref{fig:d2kpipipi} for a full description. The dashed line here represents the partially reconstructed, but Cabibbo favoured, $\Bs\to\D \Km\pip$, and charge-conjugated, decays where the pion is not reconstructed. The favoured mode cross-feed is included in the fit, but is too small to be seen.
\label{fig:d2pikpipi}}
\end{center}
\end{figure*}

\begin{figure*}[htb]
\begin{center}
\includegraphics[width=0.48\textwidth]{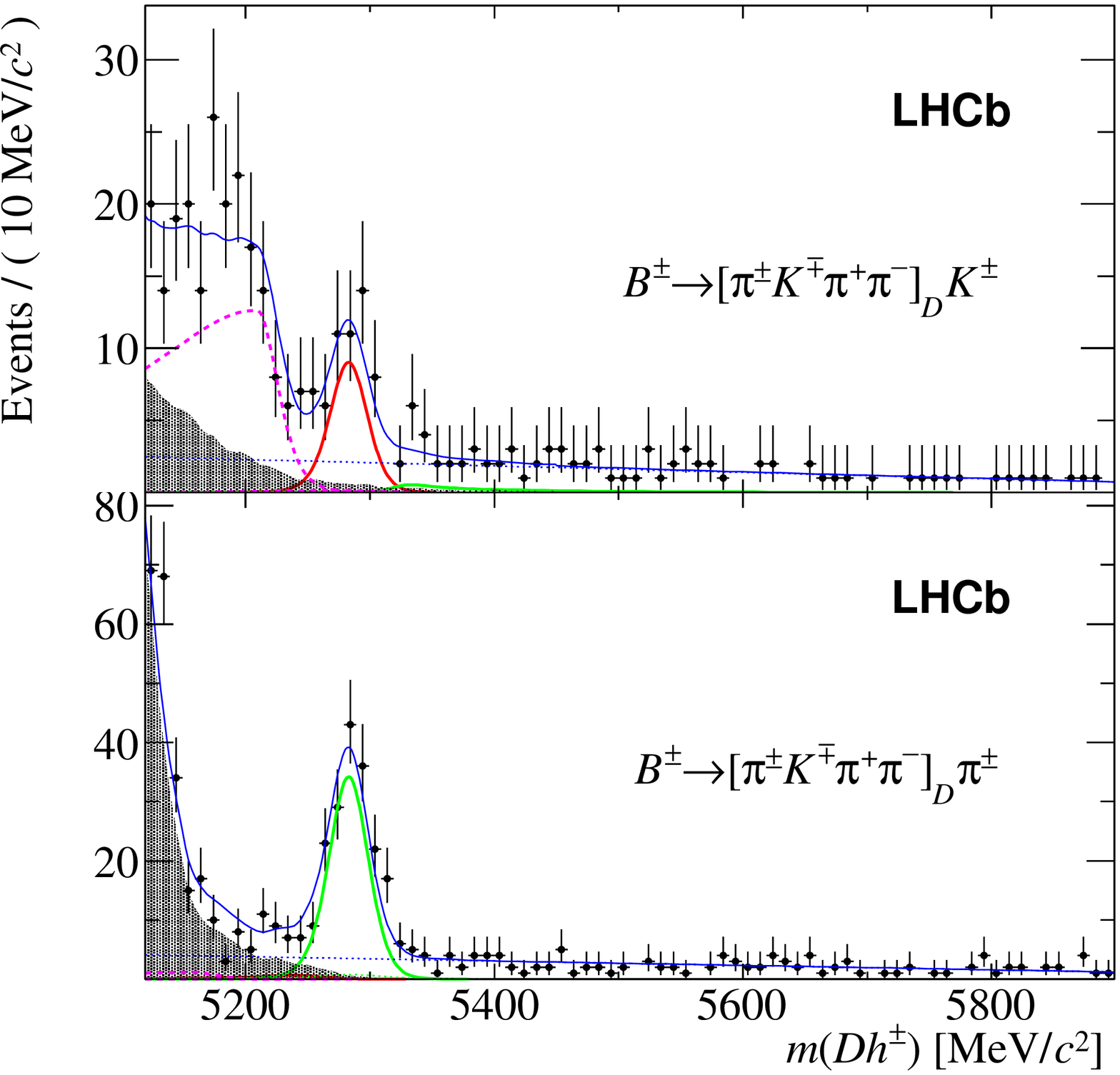}
\caption{Charge-integrated invariant mass distributions of those candidates shown in Fig.~\ref{fig:d2pikpipi} passing the $\Bpm\to[\pipm K^{\mp} \pip \pim]_Dh^{\pm}$ selection.
\label{fig:d2pikpipi_summed}}
\end{center}
\end{figure*}

The signal yields for the favoured and suppressed $\Bpm\to D h^{\pm}$ decays, after summing the events that pass and fail the bachelor PID cut, are shown in Table~\ref{tab:yields}. Their corresponding invariant mass spectra, separated by the charge of the \B candidate, are shown in Figs.~\ref{fig:d2kpipipi} and \ref{fig:d2pikpipi}. Plots of the combined \Bp and \Bm suppressed-mode mass spectra are shown in Fig.~\ref{fig:d2pikpipi_summed}.

\begin{table}[!tb]
\vspace*{-0.5cm}
\begin{center}
\begin{small}
\caption{\small Favoured ($[ K \pi \pi \pi]_{D} h$) and suppressed ($[ \pi K \pi \pi]_{D} h$) signal yields together with their corresponding statistical uncertainties. 
\label{tab:yields}}
\begin{tabular}{l|c|c}
Mode & \Bm & \Bp \\
\hline&&\\[-2.5ex]
$[ K \pi \pi \pi]_{D} \pi$  &  $20,791 \pm 232$                      &  $21,054 \pm 235$ \\
$[ K \pi \pi \pi]_{D} K$    &  $\phantom{2}1,567 \pm \phantom{0}57$  &  $\phantom{2}1,660 \pm \phantom{0}60$ \\
$[ \pi K \pi \pi]_{D} \pi$  &  $\phantom{20,7}87 \pm \phantom{0}11$  &  $\phantom{21,0}68 \pm \phantom{0}10$ \\
$[ \pi K \pi \pi]_{D} K$    &  $\phantom{20,7}11 \pm \phantom{00}5$  &  $\phantom{21,0}29 \pm \phantom{00}7$
\end{tabular}
\end{small}
\end{center}
\end{table}

The $R^{K3\pi,\pm}_{h}$ observables are related to the ratio of event yields by the relative efficiency, independent of PID effects, with which $\Bpm\to\D\Kpm$ and $\Bpm\to\D \pipm$ decays are reconstructed. This ratio is found to be 1.029 from a simulation study. A 2.4\% systematic uncertainty, based on the finite size of the simulated sample, accounts for the imperfect modelling of the relative pion and kaon absorption in the tracking material.

The fit is constructed such that the observables of interest are free parameters.  To estimate the systematic uncertainties arising from the imperfect knowledge of several of the external parameters discussed above, the fit is performed many times varying each input by its assigned error. The resulting spread (RMS) in the value of each observable is taken as the systematic uncertainty on that quantity and is summarised in Table~\ref{tab:syst}. Correlations between the uncertainties are considered negligible, so the total systematic uncertainty is the sum in quadrature of the individual components.

\begin{table}[htp]
\begin{center}
\begin{small}
\caption{Systematic uncertainties on the observables. Bachelor PID refers to the fixed efficiency for the bachelor track ${\rm DLL}_{K\pi}$ requirement determined using the \Dstarpm calibration sample. PDFs refers to the variations of the fixed shapes in the mass fit. Simulation refers to the use of simulation to estimate relative efficiencies of the signal modes, and also includes the contribution from the uncertainty in the residual background from charmless $B$ decays. $A_{\rm instr.}$ quantifies the uncertainty on the production, interaction and detection asymmetries.
\label{tab:syst}}
\begin{tabular}{l|ccccccc}
 $[\times 10^{-3}]$     & $R_{K/\pi}^{K3\pi}$  & $A_{\pi}^{K3\pi}$    & $A_{K}^{K3\pi}$    & $R_{K}^{K3\pi,-}$ &  $R_{K}^{K3\pi,+}$    & $R_{\pi}^{K3\pi,-}$  & $R_{\pi}^{K3\pi,+}$ \\ \hline
Bachelor PID   & 1.7 & 0.2 &  0.6 & 0.4 & 0.4 &  0.02  & 0.04  \\
PDFs & 1.2 & 1.3 &  4.4 &  0.7 & 0.9 &  0.09 &  0.08  \\ 
Simulation    &1.5 & 0.1 &  0.3 &  0.1 & 0.2 & 0.01  & 0.02  \\
$A_{\rm instr.}$ &0.0  &   9.9  &17.1 &  0.1  & 0.1  & 0.06  & 0.06 \\ \hline
Total  & 2.6 & 10.0 & 17.7 & 0.8 & 1.0 & 0.11 &  0.11\\
\end{tabular}
\end{small}
\end{center}
\end{table}

\section{Results and interpretation}
The results of the fit with their statistical and systematic uncertainties are
\begin{displaymath}
\begin{array}{l c r@{.}l l}
R_{K/\pi}^{K3\pi} & = &  0 & 0771\phantom{0}  \hspace{\arraycolsep} \pm \hspace{\arraycolsep} 0.0017            & \pm \hspace{\arraycolsep} 0.0026 \\[8pt]

A_K^{K3\pi}       & = & -0 & 029\phantom{00}  \hspace{\arraycolsep} \pm \hspace{\arraycolsep} 0.020\phantom{0}  & \pm \hspace{\arraycolsep} 0.018  \\[8pt]

A_{\pi}^{K3\pi}   & = & -0 & 006\phantom{00}  \hspace{\arraycolsep} \pm \hspace{\arraycolsep} 0.005\phantom{0}  & \pm \hspace{\arraycolsep} 0.010  \\[8pt]

R_{K}^{K3\pi,-}   & = &  0 & 0072\phantom{0}^{\;\;\;\; + \;\;\;\; 0.0036}_{\;\;\;\; - \;\;\;\; 0.0032}          & \pm \hspace{\arraycolsep}0.0008  \\[8pt]

R_{K}^{K3\pi,+}   & = &  0 & 0175\phantom{0}^{\;\;\;\; + \;\;\;\; 0.0043}_{\;\;\;\; - \;\;\;\; 0.0039}          & \pm \hspace{\arraycolsep}0.0010  \\[8pt]

R_{\pi}^{K3\pi,-} & = &  0 & 00417^{\;\;\;\; + \;\;\;\; 0.00054}_{\;\;\;\; - \;\;\;\; 0.00050}                  & \pm \hspace{\arraycolsep}0.00011 \\[8pt]

R_{\pi}^{K3\pi,+} & = &  0 & 00321^{\;\;\;\; + \;\;\;\; 0.00048}_{\;\;\;\; - \;\;\;\; 0.00045}                  & \pm \hspace{\arraycolsep}0.00011 \ . \\[5pt]
\end{array}
\end{displaymath}
\noindent
From these measurements, the quantities $R^{K3\pi}_{\ads (h)}$ and $A^{K3\pi}_{\ads (h)}$ can be deduced.  These are, respectively, the ratio of the suppressed to the favoured partial widths for the decays $B^\pm \to D h^\pm$, averaged over the two charges, and the \CP asymmetry of the suppressed decay mode
\begin{eqnarray*}
R^{K3\pi}_{\ads(K)}   &=& \frac{R_{K}^{K3\pi,-} + R_{K}^{K3\pi,+}}{2} =\phantom{-} 0.0124 \hspace{\arraycolsep} \pm \hspace{\arraycolsep}  0.0027 \\[8pt]
A^{K3\pi}_{\ads(K)}   &=& \frac{R_{K}^{K3\pi,-} - R_{K}^{K3\pi,+}}{R_{K}^{K3\pi,-} + R_{K}^{K3\pi,+}} = -0.42\phantom{00} \hspace{\arraycolsep} \pm \hspace{\arraycolsep}  0.22 \\[8pt]
R^{K3\pi}_{\ads(\pi)} &=& \frac{R_{\pi}^{K3\pi,-} + R_{\pi}^{K3\pi,+}}{2} = \phantom{-}  0.0037 \hspace{\arraycolsep} \pm \hspace{\arraycolsep}  0.0004 \\[8pt]
A^{K3\pi}_{\ads(\pi)} &=& \frac{R_{\pi}^{K3\pi,-} - R_{\pi}^{K3\pi,+}}{R_{\pi}^{K3\pi,-} + R_{\pi}^{K3\pi,+}} = \phantom{-} 0.13\phantom{00} \hspace{\arraycolsep} \pm \hspace{\arraycolsep} 0.10 \ .
\end{eqnarray*}
\noindent
The displayed uncertainty is the combination of statistical and systematic contributions. Correlations between systematic uncertainties are taken into account in the combination. It can be seen that the observable $A^{K3\pi}_{\ads(K)}$, which is expected to manifest significant \CP violation, differs from the \CP-conserving hypothesis by around $2\sigma$. 

A likelihood ratio test is employed to assess the significance of the suppressed \ads signal yields. This has been performed calculating the quantity $\sqrt{-2{\rm ln}\frac{\mathcal{L}_{\rm b}}{\mathcal{L}_{\rm s+b}}}$, where $\mathcal{L}_{\rm s+b}$ and $\mathcal{L}_{\rm b}$ are the maximum values of the likelihoods in the case of a signal-plus-background and background-only hypothesis, respectively. Significances of \mbox{$5.7 \sigma$} and greater than \mbox{$10\sigma$} are determined for the modes $\B^\pm \to [\pi^\pm K^\pm \pi^+\pi^-]_{D} K^\pm$ and $\B^\pm\to [\pi^\pm K^\mp \pi^+\pi^-]_{D} \pi^\pm$, respectively. The former significance is found to reduce to \mbox{$5.1 \sigma$} when the systematic uncertainties are included.

The measured observables are used to infer a confidence interval for the value of the suppressed-to-favoured $B^\pm \to DK^\pm$ amplitude ratio, $r^K_B$. 
The most probable value of $r^K_B$ is identified as that which minimises the $\chi^2$ calculated from the measured observables and their predictions for the given value of $r^K_B$.
The prediction for $R^{K3\pi,\pm}_K$ is given by Eq.~\ref{eq:rads}, and similar relations exist for the other observables.
Amongst the other parameters that determine the predicted values, $r_B^{h}$, $\delta_B^{h}$ and $\gamma$ vary freely, but all the parameters of the $D$ decay, notably the coherence factor and strong-phase difference, are constrained by the results in Ref.~\cite{NORMLOWREY}. 
Subsequently, the evolution of the minimum $\chi^2$ is inspected across the range $(0.0<r^K_B<0.2)$ and the difference $\Delta\chi^2$ with respect to the global minimum is calculated.
The probabilistic interpretation of the $\Delta\chi^2$ at each value of $r^K_B$ is evaluated by generating and fitting a large number ($10^7$) of pseudo-datasets around the local minimum. 
The variation of the pseudo-datasets is derived from the covariance matrix of the principal result.
At a given fixed point in the $r_B^K$ range, $a$, with a value of $\Delta\chi^2_a$ above the global minimum, the probability of obtaining the observed result is defined as the number of pseudo-experiments with $\Delta\chi^2\geq\Delta\chi^2_a$.
By this frequentist technique it is found that the result for $r^K_B$ has a  non-Gaussian uncertainty, so the ``$1\sigma$'' and ``$2\sigma$'' intervals, respectively, are given as
\begin{equation*}
r^K_B = 0.097\pm 0.011\ [68.3\%{\rm\ CL}]\,\,\, {\rm and } \,\,\, ^{+0.027}_{-0.029}\ [95.5\%{\rm\ CL}]\ . 
\end{equation*}
The measurements do not allow significant constraints to be set on the other underlying physics parameters.

\section{Conclusions}
\label{sec:conclusions}

A search has been performed for the ADS suppressed modes $B^\pm \to [\pi^\pm K^\mp \pi^+\pi^-]_D K^\pm$ and $B^\pm \to [\pi^\pm K^\mp \pi^+\pi^-]_D \pi^\pm$  using $1.0\,{\rm fb^{-1}}$ of data collected by LHCb in 2011.  First observations have been made of  both decays, with a significance of $5.1\sigma$ and greater than $10\sigma$, respectively.  Measurements have been made of the observables $R^{K3\pi}_{K/\pi}$, $A^{K3\pi}_h$ and $R_h^{K3\pi,\pm}$, as well as the derived parameters $R^{K3\pi}_{\ads(h)}$ and $A^{K3\pi}_{\ads(h)}$, which relate the partial widths of the $B^\pm \to Dh^\pm$ ($h=K,\pi$) family of decays.     From these observables it is deduced that $r_B^{K} = 0.097 \pm 0.011$,  where $r_B^K$ is the ratio of the absolute values of the suppressed and favoured $B^\pm \to D K^\pm$  amplitudes.   These results will improve knowledge of the Unitarity Triangle angle $\gamma$ when they are combined with other $B^\pm \to DK^\pm$ measurements  exploiting different $D$ decay modes.

\section*{Acknowledgements}

\noindent We express our gratitude to our colleagues in the CERN
accelerator departments for the excellent performance of the LHC. We
thank the technical and administrative staff at the LHCb
institutes. We acknowledge support from CERN and from the national
agencies: CAPES, CNPq, FAPERJ and FINEP (Brazil); NSFC (China);
CNRS/IN2P3 and Region Auvergne (France); BMBF, DFG, HGF and MPG
(Germany); SFI (Ireland); INFN (Italy); FOM and NWO (The Netherlands);
SCSR (Poland); ANCS/IFA (Romania); MinES, Rosatom, RFBR and NRC
``Kurchatov Institute'' (Russia); MinECo, XuntaGal and GENCAT (Spain);
SNSF and SER (Switzerland); NAS Ukraine (Ukraine); STFC (United
Kingdom); NSF (USA). We also acknowledge the support received from the
ERC under FP7. The Tier1 computing centres are supported by IN2P3
(France), KIT and BMBF (Germany), INFN (Italy), NWO and SURF (The
Netherlands), PIC (Spain), GridPP (United Kingdom). We are thankful
for the computing resources put at our disposal by Yandex LLC
(Russia), as well as to the communities behind the multiple open
source software packages that we depend on.

\addcontentsline{toc}{section}{References}
\bibliographystyle{LHCb}
\bibliography{adsglw}

\end{document}